\documentclass[pdflatex,tex,twocolumn,epjc3]{svjour3}
\RequirePackage[T1]{fontenc}
\usepackage{color}
\newcommand{\beqa}{\begin{eqnarray}}
\newcommand{\eeqa}{\end{eqnarray}}

\usepackage{float}
\usepackage{epsfig}
\usepackage{graphicx}
\usepackage{latexsym}

\usepackage{mathrsfs}
\usepackage{dcolumn}% Align table columns on decimal point
\usepackage{bm}%

\usepackage{indentfirst}
\usepackage{amssymb,dcolumn}
\usepackage{latexsym}

\usepackage{setspace}
\usepackage{graphicx,float}
\usepackage[colorlinks,linkcolor=blue,citecolor=blue,urlcolor=blue,hyperindex]{hyperref}
\usepackage{color}
\usepackage{slashed}
\usepackage{amsmath}
\usepackage{adjustbox}
\usepackage{booktabs}
\usepackage{multirow}

\journalname{Eur. Phys. J. C}
\begin{document}

\title{Cosmological constraints on two vacuum decay models}

\author{Yupeng Yang\thanksref{e1}, Yicheng Wang and Xinyi Dai}
\thankstext{e1}{e-mail: ypyang@aliyun.com}

\institute{School of Physics and Physical Engineering, Qufu Normal University, Qufu, Shandong, 273165, China}

\date{Received: date / Accepted: date}

\maketitle

\begin{abstract}
We constrain two vacuum decay models ($\Lambda(t)$CDM, proposed by the authors of~\cite{Brito:2024bhh}) utilizing the baryon acoustic oscillations (BAO) data released by 
the Dark Energy Spectroscopic Instrument (DESI), distance prior from the cosmic microwave background (CMB) observed by the Planck satellite, 
Hubble rate data obtained via the cosmic chronometers (CC) method and 
type Ia supernova (SNIa) data. The interaction terms between dark matter and dark energy are defined as 
$Q=3\varepsilon H\rho_{\Lambda}$ for model I and $3\varepsilon aH\rho_{\Lambda}$ for model II. We find that the decay parameter is constrained to be 
$\varepsilon=0.0094^{+0.0037}_{-0.0033}$ for model I and $\varepsilon=0.0119\pm{0.0045}$ for model II, 
respectively, indicating a potential interaction between dark matter and dark energy at the $2\sigma$ confidence level. 
The current Hubble parameter values are estimated to be $H_{0}=70.30\pm{0.67}$ for model I and $H_{0}=70.28\pm{0.64}$ for model II. 
These values of $H_0$ fall between those derived from the Planck and SH0ES data, suggesting that these two vacuum decay models could 
provide a potential solution to alleviate the Hubble tension problem.

\end{abstract}

\maketitle

\section{Introduction} 

In the classical standard model of cosmology, based on different astronomical observations, 
it is well known that the Universe is primarily composed of baryonic matter, dark matter, and dark energy. 
Unlike baryonic and dark matter, dark energy drives the accelerating expansion of the Universe 
and its nature remains unknown. Among the various models proposed to explain dark energy, 
the most extensively studied is the cosmological constant term $\Lambda$, which does not vary with time 
and has an equation of state parameter $w =p_{\Lambda}/{\rho_{\Lambda}} =-1$. 
A universe with the cosmological constant as dark energy is commonly referred to as the $\Lambda$CDM cosmology model, where CDM denotes cold dark matter. 

While the cosmological constant term represents the simplest approach to characterizing dark energy, it is not without its limitations and challenges
~\cite{Perivolaropoulos:2021jda}. 
Based on current observational data, the energy density associated with the cosmological constant term is estimated to be 
$\rho_{\Lambda}\sim 10^{-47}\rm GeV^{4}$~\cite{Copeland:2006wr}. 
If the cosmological constant originates from vacuum energy, quantum field theory predicts an energy density of 
$\rho_{\rm vac}\sim 10^{74}\rm GeV^{4}$, derived from the cutoff of vacuum energy at the Planck scale $m_{\rm pl}$. 
This results in a discrepancy of approximately 121 orders of magnitude between the two values. 
Even when considering the energy scale of quantum chromodynamics, the vacuum energy density is estimated to be $\rho_{\rm vac}\sim 10^{-3}\rm GeV^{4}$, 
which is about 44 orders of magnitude higher than the observed value. Another critical issue for the $\Lambda$CDM model is 
the Hubble tension~\cite{DiValentino:2021izs,Hu:2023jqc}. The Hubble constant $H_0$ can be determined through various astronomical observations. 
For instance, analyses of type Ia supernova, gravitational lensing, 
and Cepheid variable stars yield a Hubble constant of $H_{0}=73.17\pm 0.86~\rm kms^{-1}Mpc^{-1}$, 
derived from low-redshift data and largely independent of specific cosmological models~\cite{2009ApJ...699..539R,Breuval:2024lsv}. Conversely, the Hubble constant 
from the analysis of cosmic microwave background (CMB) radiation, particularly when combined with baryon acoustic oscillations (BAO) and Planck data, 
is $67.4\pm 0.5~\rm kms^{-1}Mpc^{-1}$, originating from high-redshift measurements and based on the $\Lambda$CDM cosmological model~\cite{2020A&A...641A...6P}. 
The discrepancy between these two values reaches 5.8$\sigma$, showing a significant challenge to the validity of the $\Lambda$CDM model.

Many approaches have been proposed to address the Hubble tension~\cite{DiValentino:2021izs,Hu:2023jqc}. One possibility is that there may be systematic issues with low-redshift measurements. 
Another approach involves modifying the $\Lambda$CDM universe model. For instance, kinetic dark energy models, 
where the equation of state parameter evolves over time ($w(z)$), have been extensively explored (see, e.g., Refs.~\cite{Copeland:2006wr} for a review). 
Recent BAO data released by the DESI collaboration support the idea that dark energy evolves over time~\cite{DESI:2024mwx,Colgain:2024xqj,Colgain:2024mtg}. 
In addition to dynamic dark energy models, interactions between dark matter and dark energy have also been proposed as 
potential solutions to the Hubble tension problem~\cite{Wang:2016lxa,Naidoo:2022rda,PhysRevD.94.043518,PhysRevD.96.043503,PhysRevD.96.103511,PhysRevD.98.123527,Yang:2018euj,DiValentino:2019ffd,DiValentino:2019jae,Nunes:2022bhn}. 
Early dark energy models have similarly been suggested as a resolution to this issue~\cite{Kamionkowski:2022pkx,Poulin:2018cxd}. 
Furthermore, the vacuum decay model, in which the cosmological constant term evolves over time, 
akin to an interaction between vacuum energy and dark matter, has been introduced to address related challenges
~\cite{Koussour:2024nhw,Oztas:2018zvi,Vishwakarma:2000gp,Overduin:1998zv,Azri:2014apa,Azri:2012fpi,Szydlowski:2015bwa,Bruni:2021msx,Macedo:2023zrd,Sengupta:2025jah}. 
Recently, the authors of Ref.~\cite{Brito:2024bhh} proposed two specific forms of vacuum decay models aimed at resolving the Hubble tension. 
Using Cosmic Microwave Background (CMB) and H0 data, they constrained these models and found that vacuum decays into dark matter, 
ruling out the non-interacting model at 6$\sigma$ confidence level. In this paper, we further investigate these two vacuum decay models 
using additional data sets, finding a potential interaction between dark matter and dark energy at the $2\sigma$ confidence level.

This paper is organized as follows. In Sec.~\ref{sec:basic}, we first briefly review the main contents of two vacuum decay models, 
then describe the datasets utilized and presents the final limiting results in Sec.~\ref{sec2}. The conclusions are given 
in Sec.~\ref{con}.
%%%%%%%%%%%%%%%%%%%%%%%%%%%%%%%%%%%%%%%%%%%%%%%

\section{The basic properties of vacuum decay model}
\label{sec:basic}
In this section, we briefly review the main contents of the vacuum decay model, and one can refer to, e.g., 
Refs.~\cite{Koussour:2024nhw,Oztas:2018zvi,Vishwakarma:2000gp,Overduin:1998zv,Azri:2014apa,Azri:2012fpi,Szydlowski:2015bwa,Bruni:2021msx} for more details. 
The vacuum decay model has been proposed by previous works. 
In the classical cosmology model, i.e. $\Lambda$CDM model, $\Lambda$ does not evolve with time. 
The key point of the vacuum decay model is that vacuum energy density changes over time $\Lambda(t)$.  
Within the framework of the standard cosmological model under the Friedmann-Robertson-Walker (FRW) metric, the continuity equation 
can be formulated as follows~\cite{Brito:2024bhh},

\beqa
&&\dot{\rho}_{\rm DM}+3H\rho_{\rm DM}=Q \\
&&\dot{\rho}_{\rm \Lambda}=-Q
\eeqa
where $\rho_{\rm DM}$ and $\rho_{\Lambda}$ are the energy density of dark matter and vacuum. 
$Q$ denotes the interaction between dark matter and vacuum dark energy, and $Q=0$ for $\Lambda$CDM model. 
Here, following the Ref.~\cite{Brito:2024bhh}, we adopt the interaction forms as, 

$$Q=
\begin{cases}
3\varepsilon H\rho_{\rm \Lambda}& \text{Model I}\\
3\varepsilon aH\rho_{\rm \Lambda}& \text{Model II}
\end{cases}$$
where $\varepsilon$ is a constant and $a=1/(1+z)$ is the scale factor. 
Based on these two forms of interaction, the expansion rate of the Universe, $H(z)$, can be expressed as follows, for Model I:

\beqa
\frac{H^{2}}{H^{2}_0}=\left(\Omega_{m}-\frac{\varepsilon \Omega_{\Lambda}}{1-\varepsilon}\right)(1+z)^{3}+\frac{\Omega_{\Lambda}(1+z)^{3\varepsilon}}
{1-\varepsilon}+\Omega_{r}(1+z)^{4}
\label{eq:model1}
\eeqa
and for Model II: 

\beqa
\frac{H^{2}}{H^{2}_0}=\frac{\Omega_{m}}{a^{3}}+\frac{\Omega_{\rm \Lambda}}{9\varepsilon^{3}a^{3}}\left[P(\varepsilon)
-Q(\varepsilon a)e^{3\varepsilon (1-a)}\right]+\Omega_{r}(1+z)^{4}
\label{eq:model2}
\eeqa
with $P(x)=9x^{3}+9x^{2}+6x+2$ and $Q(x)=9x^{2}+6x+2$. $\Omega_{i}\equiv 8\pi G\rho_{i}/(3H_{0}^{2})$ 
is the energy density parameter of different component at present. Note that here we consider flat $\Lambda$CDM cosmology which has a relation of 
$\Omega_{m}+\Omega_{\Lambda}+\Omega_{r}=1$.

\section{The datasets and constrains}
\label{sec2}
\subsection{Baryon acoustic oscillation}

The Dark Energy Spectroscopic Instrument (DESI) have released the first-year data (DR1), which includes six different classed of traces~\cite{DESI:2024mwx}:
the Bright Galaxy Sample (BGS,$z_{\rm eff}=0.30$), the Luminous Red Galaxy Sample (LRG, $z_{\rm eff}=0.51$ and 0.71), 
the Emission Line Galaxy Sample (ELG, $z_{\rm eff}=1.32$), 
the combined LRG and ELG Sample (LRG+ELG, $z_{\rm eff}=0.93$), 
the Quasar Sample (QSO, $z_{\rm eff}=1.49$) and the Lyman-$\alpha$ Forest Sample (Ly$\alpha$, $z_{\rm eff}=2.33$). 
The released data by the DESI collaboration are in the form of $D_{\rm M,H,V}/r_{d}$ 
and here we fix the value $r_{d}=147.09~\rm Mpc$ obtained using all CMB and CMB lensing information~\cite{2020A&A...641A...6P}. 
In the context of a homogeneous and isotropic cosmology, the transverse comoving distance, $D_{\rm M}(z)$, can be written as~\cite{2020A&A...641A...6P},

\beqa
D_{\rm M}(z) = \frac{c}{H_0}\int^{z}_{0}\frac{dz^{'}}{H(z^{'})/H_0},
\eeqa
where $c$ is the light speed and $H(z)$ is the Hubble parameter given by Eqs.~(\ref{eq:model1}) and (\ref{eq:model2}). The distance 
variable is defined as $D_{\rm H}(z)=c/H(z)$, and then the angle-averaged distance $D_{\rm V}$ can be written as 

\beqa
D_{\rm V}(z) = \left[zD_{\rm M}(z)^{2}D_{\rm H}(z)\right]^{\frac{1}{3}}
\eeqa

The $\chi^{2}$ for DESI BAO data is

\beqa
\chi^{2}_{\rm BAO}=\sum_{i}\Delta D_{i}^{T}{\rm Cov}^{-1}_{\rm BAO} \Delta D_{i},
\eeqa
where ${\rm Cov}_{\rm BAO}$ is non unit covariance matrix for the tracers of LGR,LGR+ELG, ELG and Ly$\alpha$ QSO, 
and a unit matrix for the tracers of BGS and QSO~\cite{Li:2024hrv}.

\subsection{Type Ia supernova}
For type Ia supernova (SNIa) observations, one of the main relevant parameters is the luminosity distance $d_{L}(z)$. 
For a flat $\Lambda$CDM universe, the luminosity distance can be written as 
\beqa
d_{L}(z) = (1+z)\int^{z}_{0}\frac{dz^{'}}{H(z^{'})/H_0}
\eeqa

The distance modules $\mu(z)$ is given by 

\beqa
\mu(z) = 5{\rm log}_{10}d_{L}(z)+25
\eeqa
 
For SNIa data, the Hubble constant is a nuisance parameter and here we use the methods given in, e.g., Refs.~\cite{Gong:2007se,PhysRevD.72.123519}, 
for calculating $\chi^{2}_{\rm SN}$, 

\beqa
\chi^{2}_{\rm SN}=A-\frac{B^{2}}{C}
\eeqa

where 

\beqa
&&A = \sum^{N}_{i}\frac{\left(\mu_{\rm obs}(z_i)-5{\rm log_{10}}(D_{L}(z_i))\right)^{2}}{\sigma^{2}_i} \nonumber \\
&&B = \sum^{N}_{i}\frac{\mu_{\rm obs}(z_i)-5{\rm log_{10}}(D_{L}(z_i))}{\sigma^{2}_i} \nonumber \\
&&C = \sum^{N}_{i}\frac{1}{\sigma^{2}_i}
\eeqa
where $D_{L}(z)=H_{0}/c\times d_{L}(z)$. 
We use the Pantheon sample consisting of a total of 1048 SNIa ranging of $0.01<z<2.3$~\cite{Pan-STARRS1:2017jku} for our constraints.

\subsection{Hubble rate from cosmic chronometers}

The Hubble rate can be determined utilizing the cosmic chronometer (CC) method~\cite{Jimenez:2001gg,Moresco:2022phi,Jimenez:2023flo}. 
This approach primarily focuses on measuring the differential age evolution of the Universe, $dt$, 
within a redshift interval $dz$~\cite{Jimenez:2001gg}. In practice, this is accomplished by studying carefully selected massive and passive galaxies. 
The Hubble rate using CC method can be written as~\cite{Jimenez:2001gg}

\beqa
H(z)=-\frac{1}{1+z}\frac{\Delta z}{\Delta t}
\eeqa

The $\chi^{2}$ for CC data is 

\beqa
\chi^{2}_{\rm CC}=\sum_{i}\frac{(H(z_i)-H_{\rm obs}(z_i))^{2}}{\sigma^{2}_i},
\eeqa
and we use 31 CC data points given in different Refs.~\cite{Li:2019nux,2010JCAP...02..008S,2012JCAP...08..006M,Moresco:2016mzx,Moresco:2015cya,Ratsimbazafy:2017vga,2014RAA....14.1221Z,Singirikonda:2020ieg} for our constraints. 

\subsection{Cosmic microwave background}
In general, the angular power spectrum data of CMB have been used to constraint the cosmological parameters~\cite{2020A&A...641A...6P}. 
For our purpose, instead of using the angular power spectrum data, 
here we utilize the distance prior, including the shift parameter $R$, the acoustic scale $l_{\rm A}$ 
and the baryon density $\Omega_{\rm b}h^{2}$, from the Planck 2018 data to constrain our parameters~\cite{2020A&A...641A...6P,Chen:2018dbv}. 
The shift parameter and acoustic scale can be written as~\cite{2020A&A...641A...6P,Liu:2018kjv,Xu:2016grp,Feng_2011} 

\beqa
&&R=\frac{1+z_\star}{c}D_{\rm A}(z_{\star})\sqrt{\Omega_{\rm M}H^{2}_0} \\ 
&&l_{\rm A}=(1+z_{\star})\frac{\pi D_{\rm A}(z_{\star})}{r_{s}(z_\star)}
\eeqa
here $z_\star$ is the redshift at the epoch of photon decoupling and we adopt the approximate form of~\cite{Hu:1995en} 

\beqa
z_{\star} =1048[1+0.00124(\Omega_{b}h^{2})^{-0.738}][1+g_{1}(\Omega_{m}h^{2})^{g_{2}}]
\eeqa
with 

\beqa
g_{1}=\frac{0.0738(\Omega_{b}h^{2})^{-0.238}}{1+39.5(\Omega_{b}h^{2})^{0.763}},~ 
g_{2}=\frac{0.560}{1+21.1(\Omega_{b}h^{2})^{1.81}}
\eeqa

$r_s$ is the comoving sound horizon and can be written as~\cite{DESI:2024mwx} 

\beqa
r_{s}(z)=\frac{c}{H_{0}}\int^{1/(1+z)}_{0}\frac{da}{a^{2}H(a)\sqrt{3(1+\frac{3\Omega_{b}h^{2}}{4\Omega_{\gamma}h^{2}}a)}}
\eeqa
where $(\Omega_{\gamma}h^{2})^{-1}=42000(T_{\rm CMB}/2.7{\rm K})^{-4}$ with $T_{\rm CMB}=2.7255\rm K$. The $\chi^{2}$ for CMB data is

\beqa
\chi^{2}_{\rm CMB}=\Delta X^{T}{\rm Cov}^{-1}_{\rm CMB} \Delta X
\eeqa
where $\Delta X=X-X^{\rm obs}$ is a vector with $X^{\rm obs}=(R,l_{\rm A},\Omega_{b}h^{2})$, 
and we use the values of $X^{\rm obs}$ and covariance matrix ${\rm Cov}^{-1}_{\rm CMB}$ from Planck 2018~\cite{Zhai:2018vmm}.

\subsection{Constraints on the models}
\begin{figure*}[htbp]
\centering
\begin{minipage}[t]{0.48\textwidth}
\centering
\includegraphics[width=8.5cm]{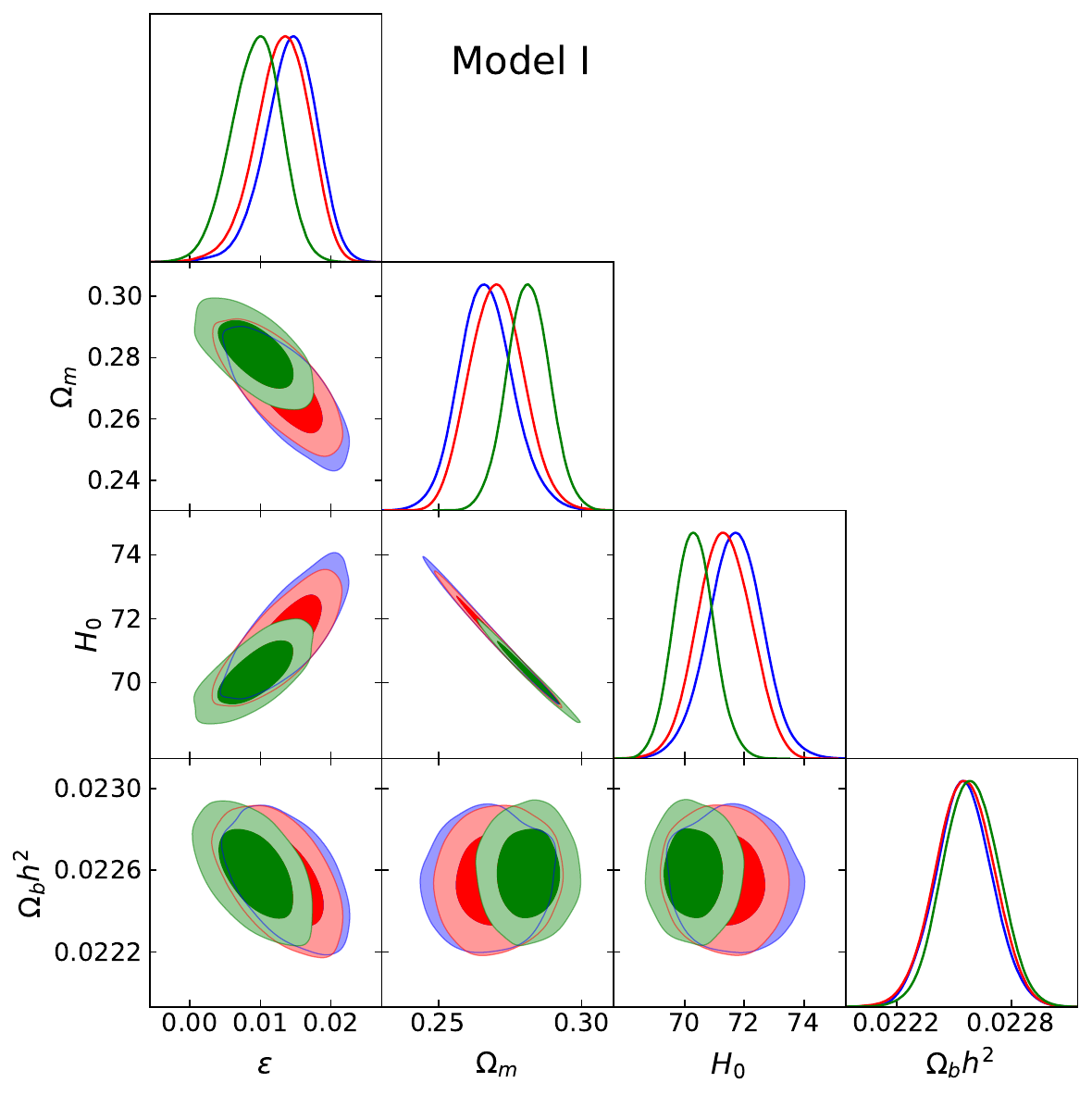}
\end{minipage}
\begin{minipage}[t]{0.48\textwidth}
\centering
\includegraphics[width=8.5cm]{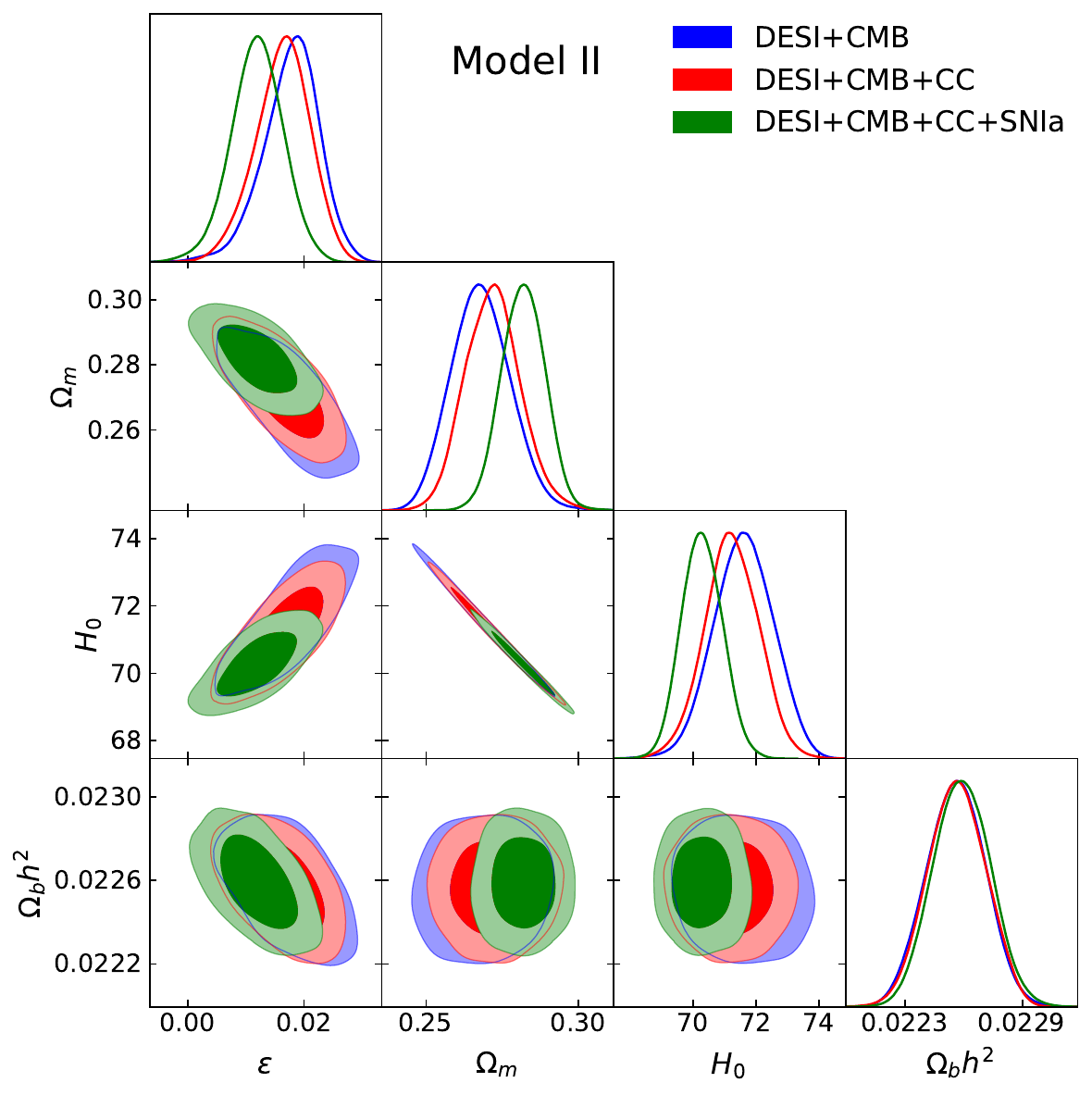}

\end{minipage}
\caption{The one-dimensional marginalized probability distribution and two-dimensional contour plots of parameters for 
model I (left) and model II (right) at the 68\% and 95\% confidence levels, 
derived using DESI+CMB, DESI+CMB+CC, and DESI+CMB+CC+SNIa data.}
\label{fig:model}
\end{figure*}

For our purpose, we use the MCMC method to find the best fit values and the posterior distributions of parameters: \{$\varepsilon$, $\Omega_{m}$, 
$H_0$, $\Omega_{b}h^2$\}. Since we will use the DESI BAO data, SNIa data, CC data and CMB data, the total $\chi^{2}$ can be written as 

\beqa
\chi^{2}_{\rm total}=\chi^{2}_{\rm BAO} + \chi^{2}_{\rm SN} + \chi^{2}_{\rm CC} + \chi^{2}_{\rm CMB}
\eeqa
The total likelihood function is $L\propto e^{-\chi^{2}_{\rm total}/2}$, and 
we employ the public code $\mathtt{emcee}$~\cite{emcee} to perform the MCMC samples. 
We set uniform prior for the parameters as $\varepsilon \in (-1,1)$, $\Omega_{m}\in (0,1)$, $H_{0}\in (50,90)$ and $\Omega_{b}h^{2} \in (0.0001,0.1)$. 
We use the public code $\mathtt{GetDist}$~\cite{getdist} to analyze the MCMC chains. 
The best-fit parameters, along with their corresponding $1\sigma$ uncertainties, are summarized in Tab.~\ref{table:cons}. 
Fig.~\ref{fig:model} shows the one-dimensional marginalized probability distributions and two-dimensional contour plots for the parameters 
derived from DESI+CMB, DESI+CMB+CC, and DESI+CMB+CC+SNIa datasets, for both Model I and Model II.

%%++++++++++++++++++++++++++ figure ++++++++++++++++++++++++++++++++++++++

%+++++++++++++++++++++++++++ figure ++++++++++++++++++++++++++++++++++++++

%****++++++++++++++++++++++++ Table I ********************************************

\begin{table*}[htb]
\caption{Constraints on the parameters of two vacuum decay models with $1\sigma$ uncertainties using various observational data combinations:
DESI+CMB, DESI+CMB+CC and DESI+CMB+CC+SNIa.}
\label{tab_ann}
\begin{center}     
%\begin{ruledtabular}  
\begin{tabular}{lllll}
\hline \hline
Model/Dataset & $\varepsilon$& $\Omega_{m}$ & $H_0$ &$\Omega_{b}h^{2}$\\

\noalign{\smallskip}\hline\noalign{\smallskip}
{\bf Model I} \\
DESI+CMB   &$0.0142^{+0.0039}_{-0.0032}$ &$0.2667^{+0.0087}_{-0.0099}$  &$71.69\pm{0.92}$ &$0.02255\pm{0.00015}$   \\
               
DESI+CMB+CC  &$0.0131^{+0.0041}_{-0.0034}$ &$0.2704\pm{0.0093}$  &$71.33\pm{0.89}$ &$0.02256\pm{0.00015}$  \\
             
DESI+CMB+CC+SNIa  &$0.0094^{+0.0037}_{-0.0033}$ &$0.2813\pm{0.0074}$  &$70.30\pm{0.67}$ &$0.02259\pm{0.00014}$  \\
\noalign{\smallskip}\hline\noalign{\smallskip}
{\bf Model II} \\
DESI+CMB   &$0.0178^{+0.0052}_{-0.0041}$ &$0.2680^{+0.0092}_{-0.0100}$  &$71.60\pm{0.95}$ &$0.02256\pm{0.00015}$   \\
               
DESI+CMB+CC  &$0.0163^{+0.0051}_{-0.0042}$ &$0.2719\pm{0.0091}$  &$71.22\pm{0.87}$ &$0.02256\pm{0.00015}$  \\
            
DESI+CMB+CC+SNIa  &$0.0119\pm{0.0045}$ &$0.2819\pm{0.0071}$ &$70.28\pm{0.64} $&$0.02259\pm{0.00015}$ \\
\hline \hline

\end{tabular}
%\end{ruledtabular}
\end{center}
\label{table:cons}
\end{table*}

%+++++++++++++++++++++++ Table I ****************************************

We have found that the best-fit values of the decay parameter are 
$\varepsilon=0.0094^{+0.0037}_{-0.0033}$ (1$\sigma$) for model I and $\varepsilon=0.0119\pm{0.0045}$ 
(1$\sigma$) for model II, respectively. Compared to the results presented in previous work~\cite{Brito:2024bhh}, 
our best-fit values for the decay parameter are smaller. 
Furthermore, we find that the noninteracting model is excluded at $2.8\sigma$ (formodel I) and $2.6\sigma$ (formodel II) 
confidence levels, which contracts with the $7.3\sigma$ (for model I) and $6\sigma$ (formodel II) confidence levels reported in Refs.~\cite{Brito:2024bhh}. 
Additionally, we have estimated the current Hubble constant to be $H_{0} = 70.30~\rm kms^{-1}Mpc^{-1}$ (for model I) and 
$H_{0} = 70.28~\rm kms^{-1}Mpc^{-1}$ (for model II), 
which are both lower than the values of $H_{0} = 73.1~\rm kms^{-1}Mpc^{-1}$ reported in Refs.~\cite{Brito:2024bhh}.

\section{Conclusions}
\label{con}

We have constrained two vacuum decay models, $\Lambda(t)$CDM, utilizing datasets from DESI BAO, CMB, CC and SNIa. 
The interaction forms of dark energy and dark matter are defined as $Q=3\varepsilon H\rho_{\Lambda}$ for model I and $3\varepsilon aH\rho_{\Lambda}$ for model II. 
We have found that the best-fit values of the decay parameter are 
$\varepsilon=0.0094^{+0.0037}_{-0.0033}$ (1$\sigma$) for model I and $\varepsilon=0.0119\pm{0.0045}$ 
(1$\sigma$) for model II, respectively. This indicates that the noninteracting model is excluded at $2.8\sigma$ (for model I) and $2.6\sigma$ (for model II) 
confidence levels. We have estimated the current Hubble constant to be $H_{0} = 70.30~\rm kms^{-1}Mpc^{-1}$ (for model I) and 
$H_{0} = 70.28~\rm kms^{-1}Mpc^{-1}$ (for model II), suggesting that these two vacuum decay models could 
provide a potential solution to alleviate the Hubble tension problem.

\section{Acknowledgements}
We thank Fengquan Wu(NAOC), Xin Zhang(NEU), Yan Gong(NAOC), Lei Feng(PMO), Shuangxi Yi(QFNU), Yankun Qu(QFNU) and Fayin Wang(NJU) for their helpful suggestions and comments. 
This work is supported by the Shandong Provincial Natural Science Foundation (Grant Nos. ZR2021MA021). 

\newcommand{\bibcommenthead}{}
\bibliographystyle{sn-aps}
\bibliography{ref}

\end{document}